\documentclass[11pt]{article} 

\usepackage{jheppub} 

\usepackage[T1]{fontenc} 

\usepackage{graphicx} 
\usepackage[dvipsnames]{xcolor}

\usepackage{booktabs} 
\usepackage{array} 
\usepackage{hyperref}
\usepackage{tikz}
\usetikzlibrary{decorations.pathmorphing}
\usetikzlibrary{decorations.markings}
\usepackage{verbatim} 
\usepackage{amsmath}
\usepackage{mathtools}

\usepackage{amssymb}
\usepackage{slashed}
\newcommand{\SLH}[1]{\left|#1\right\rangle}
\newcommand{\SRH}[1]{\left|#1\right]}

\title{Prospects for direct CP tests of $hqq$ interactions}

\author[a]{Rodrigo Alonso,}

\author[b]{Cristofero Fraser-Taliente,}

\author[b]{Chris Hays,} 

\author[a]{Michael Spannowsky}

\affiliation[a]{Institute for Particle Physics Phenomenology, Department of Physics, Durham University, South Rd, 
Durham, UK}

\affiliation[b]{Department of Physics, Oxford University, Keble Rd, Oxford, UK}

\abstract{ 
We study the prospects for probing the CP structure of $hqq$ interactions using the decays of the lightest baryon 
$\Lambda_q$ formed in the quark's hadronization.  The low yields of reconstructible events make it unlikely for  
tests to be performed with the next generation of colliders.  In $h\to b\bar b\to \Lambda_b \bar\Lambda_b$ decays 
a CP-sensitive distribution could be measured with a high-luminosity $e^+ e^-$ collider, while in both 
$h\to b\bar b\to \Lambda_b \bar\Lambda_b$ and $h\to c\bar c\to \Lambda_c \bar\Lambda_c$ decays such a distribution 
could be measured with a very high luminosity $\mu^+ \mu^-$ collider.  However, we find that only the $\mu^+ \mu^-$ 
collider can produce enough $h\to b\bar b\to \Lambda_b \bar\Lambda_b$ decays to probe a physical CP asymmetry in the 
$hbb$ vertex.
}

\begin{document}
	\maketitle

	\section{Introduction}
The Higgs boson of the Standard Model ($h$) is a CP-even state, and all CP-sensitive measurements are 
consistent with this hypothesis.  These measurements include direct tests of the CP structure of $hVV$ 
($V=g,W,Z$)~\cite{atlashvv,cmshvv}, $h\tau\tau$~\cite{cmshtautau}, and $htt$~\cite{atlastth,cmstth} 
couplings.  Tests of the Higgs boson couplings to other fermions are more challenging, due to the 
limited measurability of the fermion polarization.  They however provide unique sensitivity to 
sources of new physics, and thus merit investigation.  In addition, methods for testing the 
CP structure of Higgs-boson interactions could be applicable to any new (pseudo)scalar that may 
be discovered.

The CP structure of the $hqq$ vertex affects the polarizations of the quark and anti-quark in the 
$h \rightarrow q\bar{q}$ decay.  For $b$ and $c$ quarks we can take $\Lambda_{QCD}/m_Q\to 0$ and use 
heavy-quark effective theory to predict the transfer of the quark spin to the hadron, see 
e.g.~\cite{Manohar:2000dt}.  In the majority of cases this information is lost in the incoherent sum over 
spin states in hadronization and decay due to parity conservation in QCD and QED.  For example, the lowest 
mass pseudoscalar mesons ($P_q$) have zero spin, so the spin information is lost in the hadronizaton process.  
The spin-1 vector mesons ($P_q^*$) preserve polarization information but it is subsequently lost in the strong 
decay $P_q^* \to P_q \pi$~\cite{Falk1994}.  Vector-meson decay to polarized photons ($P_q^* \to P_q \gamma$) 
preserves spin information, but it is not expected to be observable in foreseen experiments.\footnote{To 
measure this polarization one would need to measure the charged tracks from a photon that converts to 
$e^+ e^-$ in the detector~\cite{conversions}.  The fraction of converting photons is expected to be a few percent, and the 
$e^+$ and $e^-$ have very low momentum due to the small mass difference between the vector and scalar mesons.}  
The measurement of heavy-quark polarization therefore relies on its hadronization to the lightest flavoured 
baryons ($\Lambda_q$), which decay weakly and preserve the original quark spin in the infinite-mass limit. 
The $\Lambda_b$ baryon has previously been used in polarization-sensitive studies of the $Zbb$ coupling at 
LEP~\cite{BUSKULIC1996,Abbiendi1998,DelphiPol}, and has been suggested for $b$-polarization studies at the 
LHC~\cite{Galanti2015}.  The primary limitations in an analysis of the CP structure of the $hqq$ 
vertex are the low rate of baryon production ($\approx 8\%$ of all heavy-flavour hadrons) and the 
small fraction of reconstructible baryon decays.  

In the following we first summarize the propagation of spin information from the Higgs boson decay to the 
observable final state, in order to provide a self-contained overview of the process. For this study we 
employ the massive spinor helicity formalism~\cite{Kleiss:1985yh,Kleiss:1988xr,Arkani-Hamed:2017jhn}, which 
allows the explicit tracing of spin correlations as well as the reduction of spin $\geq 1$ representations in 
terms of spin $1/2$ tensor products. For completeness we consider both $\Lambda_q$ and (vector) 
meson $P_q^{(*)}$ production, and show how the formalism provides a simple derivation of the loss of 
spin information for the latter.  We then estimate the rates of production and decay at various future 
colliders, from which we conclude that CP-sensitive observables may be measured at a $e^+ e^-$ collider 
with a luminosity upgrade or at a high-energy $\mu^+ \mu^-$ collider, but only the $\mu^+ \mu^-$ collider 
can produce enough Higgs boson decays to constrain the CP structure of the $hbb$ vertex.  

\section{The decay process}

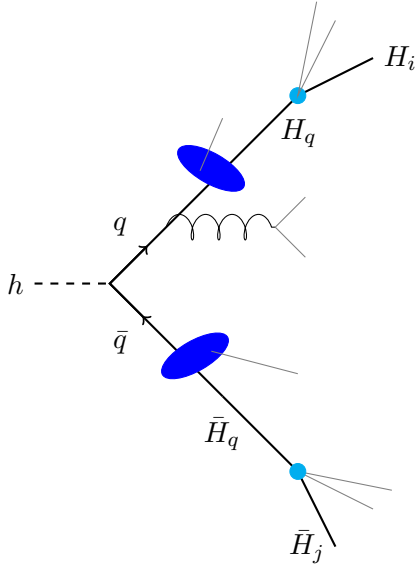
\begin{figure}[hbtp]
\centering
\begin{tikzpicture}
\draw[thick, dashed] (-1,0) node [left] {$h$} -- (0,0);
\draw [>->,thick] (.5,-.5) node [anchor=north east] {$\bar q\,\,$} -- (0,0) -- (0.5,.5) node [anchor=south east] {$q\,\,$};
\draw [decorate,decoration={coil, amplitude=1.7mm}] (0.75,0.75) -- (2.2,0.75);
\draw [gray] (2.6,0.35) -- (2.2,0.75)--(2.6,1.15);
\draw [thick] (3/2,-3/2) -- (0,0) -- (3/2,3/2);
\draw [thick] (3/2,-3/2)-- (2,-2) node [left] {$\bar H_q \,\,$} -- (5/2,-5/2) -- (3,-7/2) node[left] {$\bar{H}_j$};
\draw [gray] (3.75,-2.75) -- (5/2,-5/2) -- (7/2,-3);
\draw [thick] (3/2,3/2)-- (2,2) node [right] {$\, \, H_q \,\,$} -- (5/2,5/2) -- (7/2,3) node[right] {$H_i$};
\filldraw [Cyan] (5/2,5/2) circle (3pt); 
\filldraw [Cyan] (5/2,-5/2) circle (3pt); 
\draw [gray] (2.75,3.75) -- (5/2,5/2) -- (3,7/2);
\filldraw [rotate=30,blue] (0.5,-1.4) ellipse (0.5 and 0.2);
\draw [gray] (1.35,-.9)-- (2.5,-1.2);
\filldraw [rotate=-30,blue] (.4,2) ellipse (0.5 and 0.2);
\draw [gray] (1.2,1.5) -- (1.5,2.2);
\end{tikzpicture}
\caption{The three stages of the decay process: (1) Higgs boson decay ($h\to q\bar q$); (2) hadronization ($q\to H_q$); 
and (3) hadron decay ($H_q \to H_i X$).} 
\label{fig1}
\end{figure}

The progression from the Higgs boson decay to the measured hadron(s) is a three-stage process 
(Fig.~\ref{fig1}), each governed by different forces and scales: 
	\begin{enumerate}
		\item [(i)] 
	the {\it elementary vertex} $h\to q\bar q$ and the subsequent perturbative QCD radiation; 
	\item[(ii)] the {\it hadronization} process, in which the transfer of the (anti-)quark's polarization 
to a hadron $H_q$ ($\bar H_q$) is described qualitatively; 
		\item[(iii)] the {\it decays of $H_q$ and $\bar H_q$}, where the distributions of the decay products 
	contain information about the original process.
\end{enumerate}

We assume that any new physics dominantly affects the $hqq$ vertex, as this is the least explored part of the 
chain---subsequent processes have been studied experimentally and show no evidence of deviations from the SM.
The hard process produces a back-to-back $q\bar q$ pair in the $h$ rest frame with anti-aligned spins due to 
angular momentum conservation.  In terms of helicity-state amplitudes, the non-vanishing elements are the 
combinations $(++)$ or $(--)$.  As Fig.~\ref{fig2} shows, a CP transformation takes one amplitude into another, 
$(--)\leftrightarrow (++)$, so to test CP invariance one compares the two amplitudes.  They cannot differ 
arbitrarily, as unitarity 
requires that they have the same modulus and hence can only differ in their phases. Furthermore these phases 
change with a rotation around the axis of $q\bar q$ emission, as can be seen by acting with such a transformation 
on the fermions. This suggests a relationship between CP violation and the relative azimuthal angle (i.e. the 
angle between the decay planes) of a final-state particle from the decay of $H_q$ and another from $\bar H_q$. 
The particles in the final state selected for our distributions are referred to as polarimeters, and are 
chosen to balance theoretical sensitivity and experimental reconstructibility.

\begin{figure}[b]
	\begin{tikzpicture}
	\draw (-7.4,0) node {$ \mathcal C\times$};
	\draw (-6.5, 0) node {$\mathcal P\,\times $};
		\draw (-7,0) node {$\Bigg[$};
	\draw (-6,0) node {$\Bigg($};
 \filldraw (-4,0) circle (3pt);
 \draw [thick,>->] (-5,0)-- (-3,0);
 \draw [thick] (-5.75,0) node[anchor=south west] {$\bar q$}-- (-2.25,0) node[anchor=south east]{$q$};
 \draw (-2,0) node {$\Bigg)$};
 \draw [blue, ->] (-5.25,-0.5)--(-4.75,-0.5); 
 \draw [blue, <-] (-3.25,-0.5) -- (-2.75,-0.5);
	\end{tikzpicture} 
	\begin{tikzpicture}
	\draw (-6.25,0) node {$=$};
	\filldraw (-4,0) circle (3pt);
	\draw [thick,<-<] (-5,0)-- (-3,0);
	\draw [thick] (-5.75,0) node[anchor=south west] {$q$}-- (-2.25,0) node[anchor=south east]{$\bar q$};
	\draw (-2,0) node {$\Bigg]$};
	\draw [blue, <-] (-5.25,-0.5)--(-4.75,-0.5); 
	\draw [blue, ->] (-3.25,-0.5) -- (-2.75,-0.5);
	\end{tikzpicture}
	\begin{tikzpicture}
	\draw (-6.25,0) node {$=$};
	\filldraw (-4,0) circle (3pt);
	\draw [thick,>->] (-5,0)-- (-3,0);
	\draw [thick] (-5.75,0) node[anchor=south west] {$\bar q$}-- (-2.25,0) node[anchor=south east]{$q$};
	\draw [blue, <-] (-5.25,-0.5)--(-4.75,-0.5); 
	\draw [blue, ->] (-3.25,-0.5) -- (-2.75,-0.5);
		\draw (-2.2,0) node {$\Bigg|$};
	\end{tikzpicture}
	\caption{Parity and charge transformations on the elementary $(--)$ amplitude.  The lines on the arrows 
represent fermion flow, and the arrows below the lines represent the spin direction. \label{fig2}}
\end{figure}

Throughout the process of $q\bar{q}$ production, hadronization, and decay, sensitivity to CP violation will be 
preserved if the state is a coherent superposition of the two helicity amplitudes.  In the following we investigate 
each factorized amplitude, $\mathcal A_{h\to \bar qq}$,~$\mathcal A_{q\rightarrow H_q}$, 
and $\mathcal A_{H_q\to H_i X}$, which enter the final amplitude 
	\begin{align}
	\mathcal A= \mathcal A_{\bar H_q\to \bar H_{j}\bar X}\left[\mathcal A_{\bar q\to \bar H_q} 
\left(\mathcal A_{h\to \bar qq}\right)\mathcal A_{q \to H_q} \right]\mathcal A_{H_q\to H_i  X},
	\end{align}
with matrix notation and implicit spin indices for the different helicity amplitudes. One of the advantages of the 
helicity amplitude formalism is that 
Lorentz invariance allows us to evaluate each sub-amplitude in the most convenient frame, provided the sum over spin 
states is performed consistently. Representations of the Lorentz group $SO(1,3)\sim SU(2)_L\times SU(2)_R$ are 
given in terms of the irreducible left- and right-handed representations, which are in the irreducible representation 
of the little group for a massive particle, $SO(3)\sim SU(2)_{LG}$.  Our notation is 
\begin{align}
&\begin{array}{c|ccc}
& SU(2)_L & SU(2)_R & SU(2)_{LG}\\ \hline 
{}^{\dot\alpha}\SRH{p^I} &0 &{\bf 1/2}  &{\bf 1/2} \\
{}_\alpha\SLH{p^I} &{\bf 1/2} & 0 &{\bf 1/2} \\  
\end{array}
 & &
	\SLH{p^I}[p_I|= p_\mu\sigma^\mu
 && \sigma^\mu_{\alpha\dot\alpha}=(1,\sigma^i)
 \label{eq:spnrs}
\end{align}
with $\alpha, \dot \alpha$ indices on the Lorentz group, $I$ a little group index and $\sigma^i$ the Pauli matrices. 
The metric for the above indices is the antisymmetric rank-2 tensor $\epsilon$.  The chiral spinors and conventions 
are given in Appendix \ref{AppA}. 


	\subsection{Elementary decay}
	At the elementary particle level the Higgs boson decay can be derived from the SM $hqq$ vertex and an 
additional dimension-6 operator,
	\begin{align}
    S_{h q\bar q}=& -\int d^4x\left(y_q \bar{q}_L H q_R+ \frac{c_q}{\Lambda^2}H^\dagger H\bar{q}_L H q_R+h.c.\right)
 \label{eq:yukawa} \\
    =&-\frac{y_q}{\sqrt{2}}\int d^4x \left[\left(1+\frac{3{\rm Re}(c_q)v^2}{2y_q \Lambda^2}\right)h\bar q q+
       \frac{3{\rm Im}(c_q)v^2}{2y_q \Lambda^2}h\bar q \gamma_5 q +\dots \right],
        \end{align} 
where $v=246$~GeV.  The three-point amplitude is
	\begin{align}\label{SpPV}
	\left(\mathcal A_{h\to {\bar q}q}\right)_{JI}&=\frac{m_{q}}{v}\left(\zeta [ \bar q_J q_I ] + 
          \zeta^*\left\langle \bar q_J q_I \right\rangle\right),&
	\zeta =&1+\frac{c_q v^2}{y_q \Lambda^2},
	\end{align}
where the Kronecker delta for colour is omitted, spinors are as given in eq.~(\ref{eq:spnrs}), and other conventions 
are defined in Appendix~\ref{AppA}.

The above equation can be taken as the starting point as it is the most general three point amplitude and indeed 
extends beyond the SMEFT case, also comprising HEFT where the expansion around $\zeta=1$ need not be a good one. In 
either case corrections to the quark mass will also be present, such as those implied by eq.~(\ref{eq:yukawa}),
$m_q = (y_q v + c_q v^2/2\Lambda^2)/\sqrt{2}$, and accounted for in eq.~(\ref{SpPV}).

Taking general expressions for spinors $\SLH{q^I(p)}=\sqrt{\sigma\cdot p} \,\chi^I$ and 
$\SRH{q^I(p)}=\sqrt{\bar \sigma\cdot p} \,\chi^I$, where $\bar\sigma^\mu=(1,-\sigma^i)$ and $\chi^I$ are the two 
eigenvectors of spin ($\vec S\cdot\vec\sigma$), one can see that a parity transformation $P$ swaps angle and square 
brackets, while charge conjugation $C$ exchanges $q\leftrightarrow \bar q$. 
$CP$ conservation would then be satisfied for $\zeta=\zeta^*$.  Unitarity requires the coefficients to be complex 
conjugates of each other, so the only possible difference is a phase.

The amplitude in matrix notation in the Higgs boson rest frame is 
\begin{align}\label{eq:PV}
\mathcal A_{h\to \bar q_J q_I}&
=\frac{m_q}{v}M_h\left(\begin{array}{cc} \zeta&0\\ 0 &-\zeta^* 
\end{array}\right)\,,
\end{align}
with $M_h=125$~GeV.


	\subsection{Hadronization} 

To model the polarization through the hadronization process, we need a qualitative description of the momentum and 
spin of the heavy-flavour hadron $H_q$.  Measurements from LEP indicate that $H_q$ takes $\approx 70\%$ of the energy 
of the quark~\cite{delphibfrag}, so we assume that the hadron is produced collinearly with the quark.  We take the 
spin of the heavy quark to be unperturbed by hadronization, as given by the leading-order approximation in HQET.  In 
order to describe the process at the amplitude level we introduce the spin of the light degrees of freedom $S_\ell$, 
which combines with the spin of the heavy quark.  The amplitude of the combination of spins that produces a given 
hadron polarization is determined by Clebsch-Gordan coefficients.

We can derive Clebsch-Gordan coefficients for the projection of $J_1\otimes J_2$ onto spin $J$ states with Lorentz-invariant 
contractions of $2J_1$ $J_1$-spinors, $2J_2$ $J_2$-spinors, and $2J$ $J$-spinors, albeit all in the rest frame. We can boost 
to obtain the coefficients in another frame, but then all spinors have the same momentum. 
This implies a degeneracy in the expression for the Clebsch-Gordan coefficients in terms of spinors.  This is partially solved 
in our case by noting that hadronization must respect parity, so we take  
\begin{align}
\mathcal A_{q\to \Lambda_q}=&\frac{a_{\Lambda_q}}{2m_q}\left(\langle  q^I\Lambda_{q,K} \rangle+[\Lambda_{q,K}q^I]\right)\simeq a_{\Lambda_q}\delta^{I}_K\\
\label{CGpS}
\mathcal A_{q\to P_q}=&\frac{a_{P_q}}{2m_q}\left(\langle q^I  S_{\ell}^L \rangle+[S_{\ell}^L q^I]\right)\\
\mathcal A_{q\to P_q^*}=&\frac{a_{P_{q}^*}}{2m_q^2}\left([q^I P_{q,K_1}^*]\langle P_{q,K_2}^* S_\ell^L\rangle+[S_\ell^L P_{q,K_1}^*]\langle P_{q,K_2}^* q^{I}\rangle\right),
\end{align}
where the $K_1,K_2$ indices are symmetrized, and in line with HQET we have approximated the hadron mass to be that of 
the heavy quark. The use of spinor notation for hadronization will clarify subsequent derivations, such as the loss 
of spin information for QED and QCD decays.  We note that in this notation amplitudes differ not only by an overall 
scaling and complex phase, but also by a factor of $\sqrt{2}$ if any $J_i\geq 1$. This follows from writing explicitly 
a sum over an intermediate resonance, e.g. for spin 1:
\begin{align}
\mathcal A_{K_1,K_2}\mathcal B^{K_1,K_2}=\mathcal A_{K_1,K_2}\mathcal B_{K_1',K_2'}\varepsilon^{K_1,K_1'}\varepsilon^{K_2,K_2'}= \mathcal A_{++}\mathcal B_{--}+\mathcal A_{--}\mathcal B_{++}-2\mathcal A_{+-}\mathcal B_{+-}\,.
\end{align}
Such factors of $\sqrt{2}$ do not however appear in our final results, since we sum over little group indices.



One can combine the hard process and hadronization to obtain an amplitude with indices for the hadron spin 
and the light QCD degrees of freedom $S_\ell$ by summing over the spin indices for the internal quarks $q$ 
and using the relations given in Appendix~\ref{AppA}. The case of a $\Lambda_q$ baryon is particularly 
simple with our approximations:
\begin{align}
\mathcal A_{h\to \Lambda_q \bar \Lambda_q X}=\mathcal A_{\bar q\to \bar \Lambda_{q}} \mathcal A_{h\to q\bar q}\mathcal A_{ q\to  \Lambda_{q}}&
=\frac{m_q M_h a_{\Lambda_q}a_{{\bar \Lambda}_q}}{v}\left(\begin{array}{cc} \zeta&0\\ 0 &- \zeta^* 
\end{array}\right).
\end{align}

	\subsection{Hadron decay}

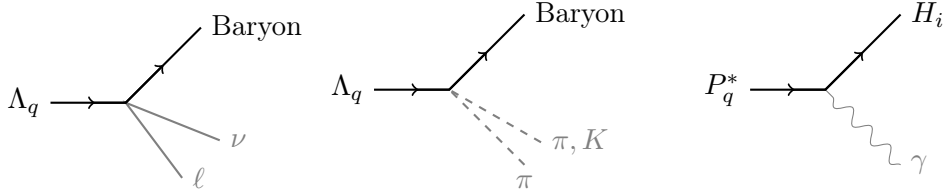
\begin{figure}[h]\centering
		\begin{tikzpicture}
		\draw[thick, gray] (1.25,-0.5) node [right] {$\nu$}-- (0,0) --(0.75,-1) node[right] {$\ell$};
		\draw [>->,thick] (-.5,0) -- (0,0) -- (0.5,.5);
		\draw [thick] (-1,0) node [left] {$\Lambda_{q}$} -- (0,0) -- (1,1) node [right] {Baryon};
		\end{tikzpicture}
	\begin{tikzpicture}
	\draw[thick, dashed,gray] (1,-1) node [below] {$\pi$}-- (0,0);
	\draw[thick, dashed,gray] (1.22,-0.7) node [right] {$\pi,K$}-- (0,0);
	\draw [>->,thick] (-.5,0) -- (0,0) -- (0.5,.5);
	\draw [thick] (-1,0) node [left] {$\Lambda_{q}$} -- (0,0) -- (1,1) node [right] {Baryon};
	\draw [gray,decorate, decoration={snake}] (6,-1) node [right] {$\gamma$}-- (5,0);
	\draw [>->,thick] (4.5,0) -- (5,0) -- (5.5,.5);
	\draw [thick] (4,0) node [left] {$P_q^*$} -- (5,0) -- (6,1) node [right] {$H_i$};
	\end{tikzpicture}
	\caption{The considered weak baryon decays (left two diagrams) and electromagnetic vector meson decay 
(right diagram).
          \label{fig3}}
\end{figure}

For hadron decay we consider the channels in Fig.~\ref{fig3}, which preserve the quark spin information and have 
sufficiently large branching ratios to be measurable in principle. For a $\Lambda_q$ baryon we consider either 
semi-leptonic decays or hadronic decays to a spin-1/2 baryon and one or two pseudoscalar mesons, while for a 
vector meson $P_q^*$ we consider the electromagnetic decay. 
 
For semi-leptonic decays the HQET limit implies that the chiral structure of weak interactions will be passed on 
to the amplitude for the $\Lambda_q$ baryon, so the amplitude is 
\begin{align}\label{slb}
\mathcal A_{\Lambda_b\to \Lambda_i\ell \bar\nu} &= \frac{4G_FV_{ib}}{\sqrt{2}} f_{1b}(q^2)  
[\Lambda_{i} |\bar\sigma_\mu| \Lambda_b \rangle J_{\ell\bar\nu}^\mu
= \frac{8G_FV_{ib}}{\sqrt{2}} f_{1b}(q^2) [\Lambda_{i} \ell] \langle \bar\nu \Lambda_b\rangle, \\\label{slc}
\mathcal A_{ \Lambda_c \to \Lambda_i \bar\ell\nu} &= \frac{4G_F V_{ci}^* }{\sqrt 2}f_{1c}(q^2) [  \Lambda_i|\bar \sigma_\mu | \Lambda_c\rangle J_{\bar\ell\nu}^\mu
=\frac{ 8 G_F V_{ci}^*}{\sqrt 2} f_{1c}(q^2) [\Lambda_i\nu]\langle  \bar\ell\Lambda_c\rangle ,
\end{align}
with $ J_{\ell\bar\nu}^\mu=[\ell| \bar\sigma^\mu|\bar \nu\rangle$, $ J_{\bar\ell\nu}^\mu=[\nu| \bar\sigma^\mu |\bar \ell \rangle$.
The decay amplitude for anti-baryons is obtained from the hermitian conjugate, which exchanges angle and square 
brackets (see Appendix~\ref{AppA}).

For a fully hadronic decay to a baryon $\Lambda_i$ and a pion, the factorized amplitude is 
\begin{align}
\mathcal A_{\Lambda_q \to\Lambda_i  \pi}= \frac{4G_FV_{iq}}{\sqrt{2}} f_{1q}[\Lambda_{i}|\bar \sigma_\mu |\Lambda_q\rangle f_{\pi}q^\mu\simeq \frac{4G_F}{\sqrt{2}} G_Ff_{1q} f_{\pi} m_{\Lambda_q} [\Lambda_{i} \Lambda_q] 
\end{align}
where the $\Lambda_i$ baryon mass is neglected in the second equality.  At this order kinematics implies that the 
spins of both $\pi$ and $\Lambda_i$ are correlated with the $\Lambda_q$ spin. For the charm quark the 
$\Lambda_c\to p K \pi$ decay is experimentally accessible.  Here the applicability of the 
factorization hypothesis is not as well motivated.  However, Bjorken has argued~\cite{Bjorken:1988ya}, and 
experimental data~\cite{Jezabek:1992ke} suggest, that the spin directions of the kaon and $\Lambda_c$ are 
anti-correlated (as is the case for $\Lambda$ in $\Lambda_c\to \Lambda \bar\ell\nu$ decays). Thus, the 
$\Lambda_c\to p K \pi$ decay is expected to be a viable mode for studies of spin correlations.

In all of the above expressions the particle whose spin is directly correlated with that of the decaying hadron 
is the one with the same angle $\langle\rangle$ or square bracket $[]$ product. This indicates the ideal polarimeter 
in theory, although in practice its detection might not be possible.
	
It is useful to collect all the above decay amplitudes into the expression 
\begin{align}\label{eqLambdaGen}
\mathcal A_{\Lambda_q \to\Lambda_i X}&=\frac{4G_FV_{iq}}{\sqrt2} [\Lambda_{i}|\bar \sigma_\mu| \Lambda_q\rangle J^\mu_X &
\left\{\begin{array}{c} J_{\pi}^\mu= f_{1q}(0) f_{\pi} \,q^\mu\\
 J_{\ell\bar\nu}^\mu=f_{1q}(q^2)[\ell| \bar\sigma^\mu|\bar \nu\rangle\\
J_{\bar\ell\nu}^\mu=f_{1q}(q^2)[\nu| \bar\sigma^\mu|\bar \ell\rangle
\end{array}\right\},
\end{align}
with the substitution $V_{iq}\to V^*_{qi}$ if the heavy quark is charm, and where we have neglected the pion mass.
 
Lastly for the electromagnetic decay of a vector meson one has: 
\begin{align}
\mathcal A_{P_q^* \to \gamma^- P_q}&=ef_\gamma \frac{\langle P_q^* \gamma\rangle^2}{m_{P_q^*}}, &
\mathcal A_{P_q^* \to \gamma^+ P_q}&=ef_\gamma\frac{[P_q^* \gamma]^2}{m_{P_q^*}}.
\end{align}
with the superscript $\pm$ representing polarization and with the same coefficient for both amplitudes, as required 
by parity.  Combining the fragmentation and decay amplitudes gives 
\begin{align}
	\mathcal{A}_{q\to P_q^*}\mathcal{A}_{P_q^*\to P_q \gamma^+}&=-ef_\gamma a_{P^*_q}m_q[S_\ell^L \gamma] [q^I \gamma ] \\
	\mathcal{A}_{q\to P_q^*}\mathcal{A}_{P_q^*\to P_q \gamma^-}&=ef_\gamma a_{P^*_q}m_q\langle S_\ell^L \gamma\rangle \langle q^I \gamma \rangle
\end{align}
where we have equated the heavy quark and hadron masses, and summed over quark spin states.  There still remains 
however the sum over the light QCD degrees of freedom $S_\ell$ in the squared amplitude; this sum together with 
the sum over photon polarizations washes away spin information, as shown in Appendix \ref{SpLss}.
Thus, one must measure the photon polarization to access the spin information of the decaying particle.

	\subsection{Total decay rate} \label{TDR}
	We finally give the rate for the full process, making explicit the little group (spin) indices that 
are usually left implicit in favour of the Lorentz group $SU(2)_L\times SU(2)_R$ structure.  The two forms are 
equivalent and here we choose the former in order to write sub-amplitudes in the frame where they are simplest. 
One has then 
	\begin{align}\nonumber
	d\Gamma = \frac{3d\Phi_{H_q} d\Phi_{\bar H_q}}{16\pi M_h (2m_{H_q})^2\Gamma_{H_q}^2 } \sum
   \Bigg[&\left(\mathcal A_{H_q \to H_i X} \mathcal  A_{H_q \to H_i X}^\dagger\right)^{\{K\}}_{\,\,\,\{I\}} 
   \left( \mathcal A_{\bar H_q \to \bar H_j X}^\dagger \mathcal A_{\bar H_q \to \bar H_j X}\right)^{\,\,\{L\}}_{\{J\}}\\
   & \times \left(\mathcal A^\dagger_{h\to \bar q  q\to  \bar H_q H_q}\right)^{\{I\},\{J\}}
 \left(\mathcal A_{h\to \bar q q \to  \bar H_q H_q}\right)_{\{L\},\{K\}}\Bigg]\\\nonumber
 =\frac{3d\Phi_{H_q} d\Phi_{\bar H_q}}{16\pi M_h (2m_{H_q})^2\Gamma_{H_q}^2 } \sum
 \Bigg[&\left( \mathcal A_{q \to H_i X} \mathcal  A_{q \to H_i X}^\dagger\right)^{K}_{\,\,I} 
 \left(\mathcal A_{\bar q \to \bar H_j X}^\dagger \mathcal A_{\bar q \to \bar H_j X} \right)^{\,\,L}_{J}\\
 & \times \left(\mathcal A^\dagger_{h\to \bar q q}\right)^{IJ}
 \left(\mathcal A_{h\to q\bar q }\right)_{LK}\Bigg],
	\end{align}
where $d\Phi$ stands for Lorentz invariant phase space and raised indices follow from the complex conjugation 
relations given in Appendix \ref{AppA}.  The indices in the first expression represent spin states of the hadron, 
and in the second expression they represent spin states of the quark-antiquark pair produced in the original 
process.  The latter provides the information contained in the original vertex: eq.~(\ref{eq:PV}) shows that 
in the Higgs rest frame the amplitude is diagonal in helicity indices and CP-violation is encoded in the 
relative phase of the two entries.  For a CP-violating effect one therefore needs interference, i.e. off-diagonal 
elements $K\neq I$, $J\neq L$ in the product of hadronization and decay amplitudes.  For QCD and QED decays these 
elements vanish and no spin information can be extracted, as shown in ref.~\cite{Falk1994}; we derive this 
result in Appendix~\ref{SpLss} using the parity of QCD and QED interactions and the spinor notation.

\begin{figure}[b]
		\begin{tikzpicture}
		\filldraw [draw=black,fill=green!20] (.2,1.8) -- (-.2,-1.8)--(-3.2,-1.8) -- (-2.8,1.8) -- (.2,1.8);
		\draw [blue] (.2,1.8) arc (85:15:0.7);
		\draw (1.2, 1.8) node {$\phi-\bar\phi$};
		\filldraw [draw=black,fill=red!10] (-.8,-1) -- (2.6,-1)--(3.8,1.3) -- (0.8,1.3) -- (-.8,-1);
		\draw [thick] (0,0) -- (2,0);\draw [thick, dashed] (2,0) -- (2.5,-.8); 
		\draw [thick] (2,0) -- (3.4,1);
		\draw (-2.5,0) arc (180:120:.6) node [anchor=east] {$\bar\theta\,\,$};
		\draw [thick] (0,0) -- (-2,0);
		\draw [thick] (-2,0) -- (-2.5,1.3); 
		\draw (1,0.3) node {$\Lambda_b$};
		\draw (-1,0.3) node {$\bar\Lambda_b$};
		\draw [thick, dashed] (-2,0) -- (-2.8,-1.5);
		\draw (2.7,0) arc (0:40:.6) node [anchor=west] {$\,\,\theta$};
		\draw [dashed] (-3,0) -- (3,0);
		\filldraw[blue!30] (0,0) circle (7pt);\draw (0,0) node {$h$};
		\end{tikzpicture}
		\begin{tikzpicture}
		\filldraw [draw=black,fill=green!20] (-.8-3.2,-1) -- (2.6-3.2,-1)--(3.8-3.2,1.3) -- (0.8-3.2,1.3) -- (-.8-3.2,-1);
		\draw [blue] (2.6-3.2,-1) arc (190:223:1.6);
		\draw (-1.1,-1.5) node {$\phi-\bar\phi$};
		\filldraw [draw=black,fill=red!10] (.2+3,1.8) -- (-.2+3,-1.8)--(-3.2+3,-1.8) -- (-2.8+3,1.8) -- (.2+3,1.8);
		\draw [thick] (0,0) -- (-2,0);\draw [thick, dashed] (-2,0) -- (-2.5,.8); 
		\draw [thick] (-2,0) -- (-3.2,-.8);
		\draw (2.5,0) arc (0:-60:.6) node [anchor=west] {$\,\,\bar\theta$};
		\draw [thick] (0,0) -- (2,0);
		\draw (1,0.3) node {$\Lambda_b$};
		\draw (-1,0.3) node {$\bar\Lambda_b$};
		\draw [thick] (2,0) -- (2.5,-1.3); 
		\draw [thick, dashed] (2,0) -- (2.8,1.5);
		\draw (-2.7,0) arc (180:220:.6) node [anchor=east] {$\theta\,\,$};
		\draw [dashed] (3,0) -- (-3,0);
		\filldraw[blue!30] (0,0) circle (7pt);\draw (0,0) node {$h$};
		\end{tikzpicture}
		\caption{CP transformation under which $\theta\leftrightarrow \bar\theta$ and 
$\phi\to\bar\phi+\pi$,$\bar\phi\to\phi+\pi$.  The polar angle is defined with respect to the axis given by the $\Lambda_b$ momentum. 
\label{fig4}}
	\end{figure}
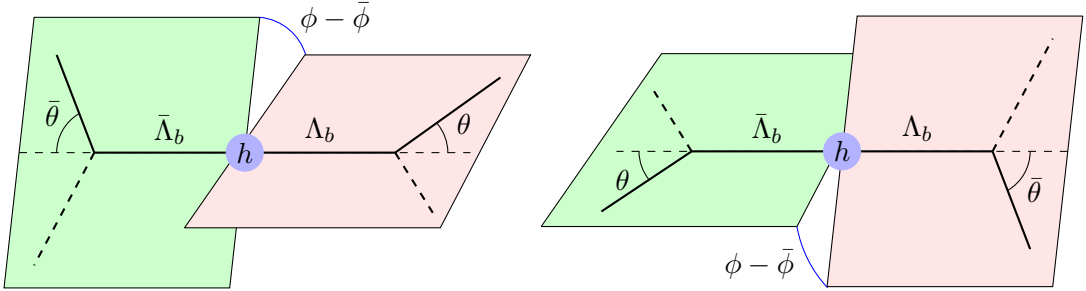

Interference is also lost when integrating over all phase space, since the rate collapses to a single sum over the 
square of amplitudes, that is
\begin{align}\nonumber
\Gamma= \sum_{I,J} \int \frac{3d\Phi_{H_q} d\Phi_{\bar H_q}}{16\pi M_h (2m_{H_q})^2\Gamma_{H_q}^2 }
\Bigg[&\left(\mathcal A_{H_q\to H_i X } \mathcal A_{H_q\to H_i X}^\dagger\right)^{\{I\}}_{\,\,\{I\}} 
\left(\mathcal A_{\bar H_q\to \bar H_j X} \mathcal A_{\bar H_q\to \bar H_j X}^\dagger\right)^{\,\,\{J\}}_{\{J\}}\\
& \times \left(\mathcal A^\dagger_{h\to  \bar q q\to  \bar H_qH_q}\right)^{\{I\}\{J\}}
\left(\mathcal A_{h\to  \bar q q\to \bar H_qH_q}\right)_{\{J\}\{I\}}\Bigg]\\
= \sum_{IJ}\Gamma_{h\to q\bar q \to H_q^{\{I\}} \bar H_q^{\{J\}}}{\rm Br}(H_q^{\{I\}}& \to H_i X){\rm Br}(\bar H_q^{\{J\}}\to \bar H_j X),
\end{align}
at which point there is no direct CP sensitivity.  We thus need a differential decay rate, which we define as a 
function of three angles: the polar angles of the two particles that we select as polarimeters, 
$\theta$ and $\bar\theta$, and their relative azimuthal angle $\delta\phi=\phi-\bar\phi$.  The former is defined in 
the rest frame of the decaying $H_q$ hadron, while the latter is invariant under a boost in the quark-antiquark 
direction.  These angles and their change under a CP transformation are shown in Fig.~\ref{fig4}. 

To summarize, $\Lambda_q$ production is the single channel preserving quark spin information, and the differential 
angular decay rate $d\Gamma/d\Omega d\bar\Omega$ is required to directly test for CP violation. It is useful to 
decompose the helicity structure of the hadron decay rate as 
\begin{align}
d\Gamma_{H_q(\hat s)}\equiv 
 \left(d\Gamma_{H_q}+d\Gamma_L \hat s_L +(d\Gamma_+ \hat s_- + c.c.)\right),
\end{align}
where $\hat s= \langle\!\langle\vec {\bf S}\rangle\!\rangle/|\langle\!\langle\vec {\bf S}\rangle\!\rangle|$ is the 
hadron spin direction, $\hat s_-=\hat s_\perp-i\hat s_T$, ${\bf S}$ is the spin operator, and 
\begin{align}\label{DPol}\nonumber
d \Gamma_{H_q}=&\frac{d\Phi}{2m_{H_q}}\frac{1}{2}\left[\left(\mathcal A_{H_q\to H_i X} \mathcal A^\dagger_{H_q\to H_i X}\right)^+_{\,\,+} +
\left(\mathcal A_{H_q \to H_i X} \mathcal A_{H_q \to H_i X}^\dagger\right)^-_{\,\,-}\right]=\frac12 {\rm Tr}\left( d\Gamma_{H_q(\hat s)}\right)\\\nonumber
d\Gamma_L
 =& \frac{d\Phi}{2m_{H_q}}\frac{1}{2}\left[\left(\mathcal A_{H_q\to H_i X} \mathcal A^\dagger_{H_q\to H_i X}\right)^+_{\,\,+} - 
\left(\mathcal A_{H_q \to H_i X} \mathcal A^\dagger_{H_q \to H_i X}\right)^-_{\,\,-}\right]=\frac12 {\rm Tr}\left({\bf S}_L d\Gamma_{H_q(\hat s)}\right) \\
d\Gamma_+=&\frac{d\Phi}{2m_{H_q}}\frac12 \left(\mathcal A_{H_q \to H_i X} \mathcal A^\dagger_{H_q \to H_i X}\right)^+_{\,\,-}=\frac14 {\rm Tr}\left(({\bf S}_\perp+i{\bf S}_T) d\Gamma_{H_q(\hat s)}\right). 
\end{align}
The perpendicular and transverse components $d\Gamma_{\perp,T}$ are twice the real or imaginary part of $d\Gamma_+$, 
\begin{align}
d\Gamma_\perp&=d \Gamma_++c.c. &
d\Gamma_T &=- id\Gamma_++c.c.
\end{align}
and we group the three terms into $d\Gamma_i=d\Gamma_\perp, d\Gamma_T, d\Gamma_L$.
For concreteness we can assign $(\perp,T,L)$ to a frame $(x,y,z)$ while bearing in mind that the polarization 
for the anti-hadron will have assignments to that frame with relative signs determined by the 
translation from helicity to spin.

   
With these definitions and the form of the elementary vertex in eq.~(\ref{eq:PV}) the rate is: 
\begin{align}\label{FDcy}
d\Gamma=\frac{3M_hd\Gamma_{H_q} d\Gamma_{\bar H_q}}{8\pi\Gamma_{H_q} \Gamma_{\bar H_q}}\frac{m_q^2|\zeta|^2}{v^2} \left(1+\frac{d\Gamma_Ld\bar{\Gamma }_L}{d\Gamma_{H_q} d\bar\Gamma_{\bar H_q}}-2\frac{\zeta}{\zeta^*}\frac{d\Gamma_+d\bar{\Gamma}_{+}}{d\Gamma_{H_q} d\bar\Gamma_{\bar H_q}}+c.c.\right)\,.
\end{align}
It is useful to rewrite this expression in terms of spin correlations and contrast with Higgs boson decay to $Z$ bosons. 
One has for Higgs bosons no net polarization for a given fermion, i.e. tracing over the spin states of other Higgs decay 
products leaves the remaining particle unpolarized, contrary to $Z$ boson decay.  To retain polarization information one 
must therefore keep track of the spin of both decay products.  Explicitly, the density matrix of elementary particles 
after Higgs boson decay is:
\begin{align}
(\rho_q)_I^{\,\,I'}\otimes (\rho_{\bar q})_J^{\,\,J'}&=\frac{\int (\mathcal A_{h\to \bar q q})_{JI}(\mathcal A_{h\to q\bar q}^\dagger)^{I'J'}d\Phi}{\int (\mathcal A_{h\to q\bar q})_{KL}(\mathcal A_{h\to q\bar q}^\dagger)^{KL}d\Phi}
\end{align}
which one can decompose in terms of spin operators as
\begin{align}
\rho_q\otimes \rho_{\bar q}&= \frac{\mbox{I\!I}}{2}\otimes \frac{\mbox{I\!I}}{2} +\frac{\mbox{I\!I}}{2}\otimes \vec{\mathcal P}_{\bar q}\cdot\vec{\bf S}+ \vec{\mathcal P}_q\cdot \vec{\bf S}\otimes\frac{\mbox{I\!I}}{2}+\mathcal T_{ij} \,{\bf S}^i \otimes{\bf S}^j.
\end{align}
It is then straightforward to check that $\vec{\mathcal P}_q=\vec{\mathcal P}_{\bar q}=0$, so that tracing over one 
of the fermions leaves the other unpolarized. The correlation between spins is
\begin{align}
\mathcal T=-\left(\begin{array}{ccc}
\cos({\rm Arg}(2\zeta)) & -\sin({\rm Arg}(2\zeta)) &0\\
\sin({\rm Arg}(2\zeta)) & \cos({\rm Arg}(2\zeta)) &0\\
0&0&1
\end{array}\right)
\end{align}
with spin as measured in the $q$ and $\bar q$ rest frames but with the axes of both frames aligned (this means 
that the explicit operator form of $\bf S$ acting on $q$ and $\bar q$ helicity states is different).

In terms of the spin correlation the rate is
\begin{align}
d\Gamma=\frac{3M_h d\Gamma_{H_q} d\Gamma_{\bar H_q}}{8\pi \Gamma_{H_q}\Gamma_{\bar H_q} }\frac{m_q^2|\zeta|^2}{v^2}
\left(1+\frac{d\Gamma^i }{d\Gamma_{\Lambda_q} }\mathcal T_{ij} \frac{\bar{d\Gamma}^j}{d\bar \Gamma_{\bar\Lambda_q}}\right).\,
\end{align} 
If the polarization fraction retained by the hadron $r_i$~\cite{Galanti2015} is less than one, this can be implemented by 
the replacement $\mathcal T_{ij}\to r_i\mathcal T_{ij} r_j$.

	\subsection{The di-baryon final state}
	We focus on the experimentally viable $\Lambda_q \times \bar\Lambda_q$ final state and neglect the effect of excited 
baryon states that slightly dilute the polarization information~\cite{Galanti2015}.  The procedure is to select a final-state 
particle to serve as the polarimeter and integrate over the phase space of the remaining particles.  The contribution of each 
secondary decay to the rate is
\begin{align}
d\Phi_{H}\left(\mathcal A_{H_q\to H_i X } \mathcal A_{H_q\to H_i X}^\dagger\right)^I_{\,\,J}&\equiv \langle \Lambda^I_q| \sigma_\mu |\Lambda_{q,J}] dK^\mu\\
d\Phi_{\bar H}\left(\mathcal A_{\bar H_q\to \bar H_j X }^\dagger \mathcal A_{\bar H_q\to \bar H_j X}\right)^{\,\,K}_{L}&\equiv -\langle \bar\Lambda_{ q,L}| \sigma_\mu |\bar\Lambda_{ q}^K] d\bar K^\mu
\end{align}
where the sign is due to the complex conjugation of spinors and little group indices, see Appendix~\ref{AppA}.
The four-vector $dK^\mu$ is proportional to a combination of the polarimeter momentum and that of the $\Lambda_q$ baryon. 
The highest sensitivity is achieved when the $p_{\Lambda_q}$ component is zero; for semileptonic decays the 
amplitudes in eqs.~(\ref{slb}) and (\ref{slc}) show that this would imply maximal sensitivity for neutrinos from $\Lambda_b$
decay and charged leptons from $\Lambda_c$ decay. 

The currents $dK^\mu$ for various decays and choices of polarimeter particle are
\begin{align}
\Lambda_b&\to\Lambda_c\pi:\quad dK_\pi^\mu = \frac{(2\pi)^2 f_{1b}^2(0)f_\pi^2}{m_{\Lambda_b}} \frac{1-\delta}{1+\delta} \, p_\pi^\mu \frac{C_{cb}d^3p_\pi}{2E_\pi(2\pi)^3}  \delta(m_{\Lambda_b}-E_{\Lambda_c}-E_\pi) \\ 
\Lambda_b&\to \Lambda_c\bar\nu\ell:\quad dK_{\ell\bar\nu}^\mu= \hat f^2_{1b}(x)\frac{(y-x)}{y^2}\left[ xp^\mu+\left(1-\delta+x-y-\frac{2x}{y}\right) p_\ell^\mu\right] \frac{ C_{cb}dx d^3p_\ell}{2E_\ell(2\pi)^3} \\
\Lambda_c&\to \Lambda\nu\bar\ell:\quad dK_{\bar\ell\nu}^\mu=\hat f^2_{1c}(x) \frac{(1-\delta-y)}{y} p_\ell^\mu \frac{C_{cs}dx d^3p_\ell}{2E_\ell (2\pi)^3}
\\
\Lambda_c&\to\Lambda \nu\bar\ell:\quad dK_{\Lambda}^\mu =  \frac{\hat f^2_{1c}( x )}{6}\left[(1-\delta-x)(p-p_{\Lambda})^\mu + x p^\mu_\Lambda\right] \frac{C_{cs}dp_{\Lambda}^3}{ 2E_k(2\pi)^3}
\end{align}
where
\begin{align}
x&=\frac{q^2}{m_{\Lambda_q}^2}\,, & y&=\frac{2p_\ell \cdot p}{m_{\Lambda_q}^2}\,,  & \delta&=\frac{m_{\Lambda_i}^2}{m_{\Lambda_j}^2}\,,& \hat f(x)&=f(m_{\Lambda_q}^2x)\,, & C_{ij}&=\frac{4|V_{ij}|^2 G_F^2 m_{\Lambda_q}^2}{\pi}\,.
\end{align}
The inclusive semi-leptonic case can be obtained from the above in the leading approximation with the substitutions $f\to 1$ 
and $\delta=m_c^2/m_b^2, m_s^2/m_c^2$. 

Given the expressions for the $\Lambda_b$ spinor in its rest frame and the decomposition in eq.~(\ref{DPol}), the 
average over $+$ and $-$ helicities selects the time-like component of $dK^\mu$.  The longitudinal contribution 
$d\Gamma_L$ selects the component along the $\bar\Lambda_b$ direction ($dK_L\propto \cos\theta$), whereas $d\Gamma_+$ 
selects normal components ($K_{\perp}+iK_T\propto\sin\theta e^{i\phi}$). The ideal polarimeter has the most sensitive 
angular distribution, which translates into the largest time-like to space-like component ratio in $dK^\mu$. This is 
conventionally encoded in the spin resolution power, $\alpha$, which here is defined as
\begin{align}
\frac{d\Gamma}{d\cos\theta_n}&= \frac{\Gamma}{2}(1+\alpha_n\cos\theta_n)& 
\alpha_n& 
=\frac{3\int \cos \theta_n d\Gamma}{\int d\Gamma},
\end{align}
with $\theta_n$ the angle between the polarimeter particle $n$ and the hadron spin direction, and the factor of 3 is 
for normalization.  The parameter $\bar \alpha$ is defined similarly for the $\bar\Lambda_b$ case.  Dividing the Higgs 
boson decay rate by $\Gamma_{q\bar q}$ gives the angular differential decay rate, 
\begin{align}
\frac{(4\pi)^2}{\Gamma_{h\to q \bar q}}
\frac{d\Gamma_{h\to \Lambda_i X \bar \Lambda_{j} X'}}{d\Omega d\bar \Omega} 
&=
{\rm Br}_{\Lambda_q \to \Lambda_i X}{\rm Br}_{\bar \Lambda_q \to \bar \Lambda_{j} X'}f_{\Lambda_q}^2
\left(1+\frac{d\Gamma^i }{d\Gamma_{\Lambda_q} }\mathcal T_{ij} \frac{\bar{d\Gamma}^j}{d\bar \Gamma_{\bar\Lambda_q}}\right)\\
&=
{\rm Br}_{\Lambda_q \to \Lambda_i X}{\rm Br}_{\bar \Lambda_q \to \bar \Lambda_{j} X'}f_{\Lambda_q}^2
\left(1+\frac{d\Gamma_Ld\bar{\Gamma}_L}{d\Gamma_{\Lambda_q} d\bar \Gamma_{\bar\Lambda_q}} -2\frac{\zeta^*}{\zeta}\frac{d\Gamma_+  d\bar {\Gamma}_{+}}{d\Gamma_{\Lambda_q} d\bar\Gamma_{\bar\Lambda_q}}+c.c.\right),
\nonumber	\\ \nonumber
&= {\rm Br}_{\Lambda_q \to \Lambda_i X}{\rm Br}_{\bar \Lambda_q \to \bar \Lambda_{j}X'} f_{\Lambda_q}^2
\left[1+\alpha_n\bar\alpha_m\left(c_\theta c_{\bar\theta} + 
{\rm Re}\left(\frac{\zeta^*}{\zeta} e^{i\delta\phi}\right)s_{\theta}s_{\bar\theta}\right)\right],
\end{align}	
where the angular coordinates $\Omega$ and $\bar \Omega$ are defined in the rest frames of $\Lambda_q$ and 
$\bar \Lambda_q$, respectively, $\delta\phi=\phi-\bar\phi$ has the orientation shown in Fig.~\ref{fig4}, 
and $f_{\Lambda_q}$ is the fragmentation ratio into $\Lambda_q$, which is $\approx 7$\% for 
$b$ quarks~\cite{Amhis:2014hma} and $\approx 8$\% for $c$ quarks~\cite{AlephLambdaC}.

The ratios $\alpha_a$ determine the quality of our polarimeter particle and are as follows (the particle used 
as the polarimeter is highlighted in blue):
\begin{center}
\begin{tabular}{c|c|c|c|c|c}
	& $\Lambda_b\to\Lambda_c {\color{blue}\ell}\nu$ & $\Lambda_b\to X_c {\color{blue}\ell}\nu$ 
        & $\Lambda_c\to\Lambda {\color{blue}\ell}\nu$ or ${\color{blue}\pi}\Lambda$ & $\Lambda_c\to X_s {\color{blue}\ell}\nu$ 
        & $\Lambda_c\to p{\color{blue}K} \pi$ or ${\color{blue}\Lambda}\ell\nu $\\ \hline
	$\alpha$& $-0.4$&  $-0.3$ & $1,1$ & $1$ & $-0.2,-0.4$\\ 
	$\bar \alpha$& $0.4$&  $0.3$ & $-1,-1$ & $-1$ & $0.2,0.4$
\end{tabular}
\end{center}
Due to the chiral structure of the weak interaction, the optimal polarimeters have opposite isospin to the decaying quark, 
as evident in the right-hand side of eqs.~(\ref{slb}) and (\ref{slc}).  Polarimeters with the same isospin retain correlations 
through momentum conservation, and for these $\alpha_a$ can be calculated from phase space integrals (see Appendix~\ref{alPWR}).  
Except for $pK\pi$, which is taken from \cite{Jezabek:1992ke}, the above values are rough estimates determined using our 
approximations and formalism, though both experimental and more precise theory inputs are available~\cite{Amhis:2014hma}.  
The relative sign $\bar\alpha=-\alpha$ can be understood intuitively from the weak group chirality: if a given decay involves 
a (massless) polarimeter particle with one helicity, the weak interactions in the conjugate process select the opposite 
helicity for the anti-particle (hence a flipped sign). 

To define a CP-odd observable we subtract the CP conjugate,
\begin{align}
\mathcal C\mathcal P(d\Gamma)(\mathcal C\mathcal P)^{-1}=d\Gamma^{CP}=d\Gamma(n\to \bar m, m\to\bar n,\theta\leftrightarrow \bar\theta, \phi\leftrightarrow\bar\phi),
\end{align}
where a bar indicates the antiparticles of the original decay product.
  
Normalizing to the total rate, the relative differential rate of CP violation is 
\begin{align}
\frac{d\Gamma- d\Gamma^{CP}}{\Gamma + \Gamma^{CP}}=\alpha_n\bar \alpha_m\sin(2{\rm Arg}\zeta)
\sin(\phi-\bar\phi)\sin\theta\sin\bar\theta\frac{ d\Omega_n d\bar\Omega_m}{(4\pi)^2}.
\end{align}
To experimentally probe CP, we integrate over the CP-sensitive observable to obtain the number of events with a given sign, 
\begin{align}
N_\pm& \equiv N [\vec{p}_m \cdot (\vec{p}_n \wedge \vec{p}_{\Lambda_b}) \gtrless 0], &
N^{CP}_\pm&\equiv N^{CP} [\vec{p}_{\bar n} \cdot (\vec{p}_{\bar m} \wedge \vec{p}_{\Lambda_b}) \gtrless 0]\,,
\end{align}
as measured in the Higgs boson rest frame.  The integrated asymmetry is then 
\begin{align}
\epsilon_{CP}^{nm} = \frac{N_+ - N_- + N_+^{CP} - N_-^{CP}}{N_+ + N_- + N_+^{CP} + N_-^{CP}}, 
\end{align}
which corresponds to 
\begin{align}
\label{CPobs}
\epsilon_{CP}^{nm}\equiv \left[\int_{\sin(\phi-\bar\phi)>0}-\int_{\sin(\phi-\bar\phi)<0}\right]   \frac{d\Gamma-d\Gamma^{CP}}{2\Gamma} = 
\frac{\pi \alpha_n \bar \alpha_m \sin(2{\rm Arg}\zeta) }{8}.
\end{align}
Given the small fragmentation into $\Lambda_q$, considering multiple decay channels can increase yields, 
although general considerations require $-1\leq\alpha\leq 1$ for each channel.

Finally, for an alternative derivation of the CP-violating effect, rather than writing amplitudes for 
every subprocess we can carry out the spin sums connecting the different vertices.  Using 
eq.~(\ref{eqLambdaGen}) and the completeness relation 
	\begin{align}
	|\Lambda_{q,K}\rangle\left(\zeta^*\langle \Lambda_q^K \bar \Lambda_{q}^L\rangle + 
\zeta[ \Lambda_q^K \bar \Lambda_{q}^L] \right)
[\bar \Lambda_{q,L}|=(\zeta^* m_{\Lambda_q} p_{\bar\Lambda_q}^\mu\sigma_\mu - 
\zeta m_{\Lambda_{q}} p_{ \Lambda_{q}}^\mu \sigma_\mu)\equiv m_{\Lambda_q}\hat\zeta_\mu\sigma^\mu\,,
	\end{align} 
the general differential decay rate reads 
	\begin{align}
	d\Gamma &= \mathcal N\, \mbox{Tr}\left(\bar \sigma_\mu \sigma_\alpha\bar\sigma_\nu\sigma_\beta \right) 
dK_{n}^\mu d\bar K^\nu_m \hat\zeta^\alpha \hat\zeta^{* \beta} 
& \left\{\begin{array}{c} K_\pi^\mu = f_\pi (p_\pi \cdot p_{\Lambda_i}) p_\pi^\mu  \\ 
K_{\nu\ell}^\mu = f(q^2)[(m_{\Lambda_q}^2-q^2)q^\mu + q^2 p^\mu_{\Lambda_i}]\end{array} \right\}
        \end{align}
 with the Lorentz structure that gives rise to CP violation appearing in 
the Levi-Civita tensor that results from tracing over  $\sigma^\mu$ matrices.

\section{Experimental sensitivity}
Given the ability to model the progression of CP information from the Higgs-boson decay to the final-state particles, we 
next investigate the potential experimental sensitivity to the CP structure of $hqq$ interactions at future colliders.  
We focus on $hbb$ and $hcc$ interactions, for which heavy-quark effective theory can be used.  Approximately one in $10^6$ 
$h\to q\bar{q}$ decays are reconstructible for a CP test because of the fragmentation of both quarks into baryons 
($\approx 1\%$), the branching ratios for both baryons to decay into measurable final states ($\approx 1\%$), and the 
reconstruction acceptance ($\approx 1\%$).  The analysis thus requires $>10^8$ Higgs-boson events, which could be produced 
by a high-luminosity $pp$, $e^+ e^-$, or $\mu^+ \mu^-$ collider. 

The proposed FCC-hh would provide an integrated luminosity of 20~ab$^{-1}$ of $\sqrt{s}= 100$~TeV $pp$ collisions in each 
of two detectors over a span of two and a half decades~\cite{fcchh}.  To study the $h\to q\bar{q}$ decay, the $h\ell\nu$ 
and $h\ell\ell$ ($\ell = e,\mu$) production processes would be needed to avoid the large QCD multijet backgrounds.  
Combining the respective cross sections of 3.4~pb and 0.75~pb~\cite{vhxsec,pdg}, and the yields of two experiments, gives 
a total of 170 million Higgs boson events.  

The FCC-ee collider would provide an integrated luminosity of 5~ab$^{-1}$ of $\sqrt{s}=240$~GeV $e^+e^-$ collisions 
at two collision points in three years of operation~\cite{fccee}.  This would be insufficient to probe CP-violation in the 
$hqq$ interaction, so we consider a high-luminosity $e^+ e^-$ collider~\cite{hlfccee} that would produce five times the 
instantaneous luminosity of the FCC-ee and would run for the same duration as the FCC-hh.  With the resulting 200~ab$^{-1}$ 
of integrated luminosity in each of four detectors, and the $e^+ e^- \to Zh$ cross section of 200~fb~\cite{fccee}, a similar 
number of Higgs bosons would be produced as for the FCC-hh $h\ell\nu$ and $h\ell\ell$ processes.

A muon collider~\cite{mucollider} offers the potential for high-energy collisions at high luminosity.  Two speculative 
scenarios under investigation~\cite{muphysics} that would produce sufficient Higgs boson yields are $\sqrt{s}=30$~TeV 
and 100 TeV $\mu^+ \mu^-$ colliders with 90 ab$^{-1}$ and 1000 ab$^{-1}$ of integrated luminosity, respectively, for 
five years of running.  The cross section for Higgs boson production through vector-boson fusion is 1.2 pb for 
$\sqrt{s}=30$~TeV~\cite{muvbf}, and extrapolating the logarithmic growth with energy to 100~TeV gives a cross section 
of 1.4 pb. The expected Higgs-boson yields are thus 110 million to 1.4 billion per detector.

	\subsection{$h\to b\bar b \to \Lambda_b \bar\Lambda_b$}

The multistage decay of $\Lambda_b \to \Lambda_c X \to \Lambda X'$ precludes the full reconstruction of the $\Lambda_b$ 
baryon with high efficiency.  In order to maximize the event yield, we consider a partial $\Lambda_b$ reconstruction that 
requires only a charged lepton (to provide polarization information) and a $\Lambda$ baryon (to provide discrimination 
from other hadrons).  This final state was successfully used in measurements of $Z\to b\bar{b}$ decays at 
LEP~\cite{AlephLambdaB,DelphiLambdaB,OpalLambdaB}.  

We estimate the expected event yields for $h\to \Lambda_b \bar \Lambda_b \to \Lambda \ell\nu \bar\Lambda \ell\nu X$ 
using the 57.5\% $h\to b\bar b$ branching fraction and the $0.65 \pm 0.08\%$ rate measured by ALEPH for 
$b \to \Lambda_b \to \Lambda \ell\nu X~(\ell=e,\mu)$~\cite{AlephLambdaB}.  Combining these factors gives a total of 
$24 \pm 3$ events per million Higgs bosons produced.  The yield is further reduced by the 64\% branching fraction of 
the observable $\Lambda \to p\pi$ decay, giving $10 \pm 1$ events with a measurable $\Lambda \ell^+ \bar \Lambda \ell^-$ 
final state per million Higgs bosons.  

Given the small event yields, sensitivity is only possible if the backgrounds are very low.  This makes the FCC-hh 
insensitive, since $h \to bb$ measurements are background-dominated in $pp$ collisions~\cite{atlashbb,cmshbb}.  The analysis 
may however be feasible in the low-background environment of an $e^+ e^-$~\cite{eehbb} or a $\mu^+ \mu^-$~\cite{muvbf} 
collider.  

We estimate the acceptance for selecting events in $e^+ e^-$~collisions using a sample of $e^+e^- \to ZH$ events 
generated with Madgraph\_aMC@NLO~\cite{madgraph} + Pythia~6.428~\cite{pythia} at leading order.  We do not include 
detector resolutions and efficiencies, which will need to be high to have CP sensitivity---the primary challenge will be 
particle identification.  We select $\Lambda_b$ decays by requiring a charged lepton and a $\Lambda$ baryon with 
respective momentum $>3$~GeV and $>4$~GeV that are within $\Delta R = \sqrt{(\Delta \eta)^2 + (\Delta \phi)^2} < 0.4$ 
of each other.  The $\Lambda$ baryon must have a decay radius $>3$~cm.  The momentum requirements are the same as 
those used in the ALEPH b-baryon lifetime measurement~\cite{AlephLambdaB}, while the decay radius requirement is looser 
in order to maintain high acceptance.  The corresponding increase in background would have to be mitigated by improved 
detector resolution or analysis algorithms.  The acceptance of the momentum requirements is 81\% per $\Lambda_b$ baryon, 
and that of the decay radius is 45\% per $\Lambda_b$.  Incorporating these acceptances, a total of 210 reconstructed 
events is expected for a high-luminosity $e^+ e^-$ collider that produces 160 million Higgs boson events. 

Sensitivity to the CP structure of the $hbb$ vertex is obtained from the signed angle between the charged leptons in the 
Higgs boson rest frame:
\begin{equation}
\sin\delta\phi = \frac{\vec{p}_{\bar\ell} \cdot (\vec{p}_\ell \times \hat{p}_{\Lambda_b})}
{\sqrt{(\vec{p}_\ell)^2 - (\vec{p}_\ell \cdot \hat{p}_{\Lambda_b})^2}
\sqrt{(\vec{p}_{\bar\ell})^2 - (\vec{p}_{\bar\ell} \cdot \hat{p}_{\Lambda_b})^2}}.
\end{equation}

\noindent
where $\hat p_{\Lambda_b}$ represents the unit vector in the direction of $\Lambda_b$.
The angle is reconstructed by taking the dijet momentum as a measure of the Higgs boson momentum, and the direction of each 
$\Lambda_b$ baryon as the direction of the corresponding jet boosted into the Higgs boson rest frame.  The difference 
between the baryon axis and the jet axis leads to a sign flip in $\delta\phi$ in 31\% of the events in the Monte Carlo.  
The asymmetry in these events will be cancelled by the same fraction of events without a sign flip, leaving 38\% of the events 
available for a measurable asymmetry.  Accounting for this reduction an asymmetry of 0.5 would be measurable at the $3\sigma$ 
level.  However, this is an order of magnitude larger than the maximum expected value of 0.04 given by eq.~(\ref{CPobs}).

Given our optimistic assumptions in the collider luminosity and detector efficiencies and resolution, any CP test of the 
$Hbb$ interaction using $e^+ e^-$ collisions appears unlikely.  With a tenfold increase in yield from a 100~TeV $\mu^+ \mu^-$ 
collider, a test of the order of the maximum expected effect could be performed.  Assuming the reconstruction acceptances 
are similar to those of an $e^+ e^-$ collider, $2\sigma$ sensitivity could be possible for an asymmetry of 0.04.  

	\subsection{$h\to c\bar c \to \Lambda_c \bar \Lambda_c$}

The $h\to c\bar c$ decay appears less promising than $h\to b\bar b$ due to its 2.9\% branching fraction.  
This is somewhat compensated by the fully reconstructible $\Lambda_c \to pK\pi$ decay.  The rate for 
$c\to \Lambda_c\to pK\pi$ has been measured by ALEPH to be 0.39\%~\cite{AlephLambdaC}, so only 0.44 
$h\to \Lambda_c\bar \Lambda_c \to pK\pi pK\pi$ events are expected per million Higgs bosons produced.  
Even if we optimistically assume a reconstruction acceptance of 25\% for each $\Lambda_c$, only 4.4 events 
would be expected at a high-luminosity $e^+e^-$ collider.  A high-luminosity $\mu^+ \mu^-$ collider could 
yield 40 events, which would provide $1.5\sigma$ sensitivity to an asymmetry of 0.25.  This is more than 
an order of magnitude larger than the maximum theoretical asymmetry of 0.02 using the $\Lambda_c \to pK\pi$ 
decay.

\section{Conclusion}

In this work we have explored the feasibility of direct CP tests in Higgs-boson decays to heavy quarks in 
future collider experiments. The effect of a CP transformation is encoded at the fundamental level in relative 
complex phases for different helicity amplitudes, which we have converted into observables in angular 
distributions of the final decay particles.  We have found that massive helicity amplitude methods are 
particularly well suited for tracing the spin correlations throughout the derivation of results.

A test of CP violation in these decays is challenging due to: $i)$ the loss of measureable spin correlations 
during hadronization in all but the $\Lambda_q$ channel, which has a $<10\%$ fragmentation fraction; and 
$ii)$ the low $\Lambda_q$ branching ratio to experimentally reconstructible final states. The Higgs boson 
yields therefore need to be very high to approach sensitivity, ${\cal{O}}(10^9)$ events, beyond the reach of 
all proposed colliders except a high-luminosity $100$~TeV muon collider.  With such a collider it may be 
possible to test maximal CP violation at the $2\sigma$ level. 

Due to the challenges in studying direct CP violation in $h\to b\bar b$ decays, it is appropriate to study 
alternative tests such as the associated production of Higgs bosons with $b$ or $c$ quarks at hadron colliders, 
or the decay of the Higgs boson to $J/\psi \gamma$.  By combining multiple approaches it may be possible to 
directly constrain CP violation in the $hbb$ and $hcc$ couplings.

	
\appendix
\section{Massive spinor conventions}\label{AppA}
The metric is taken in each $SU(2)$ space as
\begin{align}
(-\epsilon^{IJ})=\epsilon^{\alpha\beta}=\epsilon^{\dot\alpha\dot\beta}=\left(\begin{array}{cc} 0&1\\ -1&0\end{array}\right)\,,\end{align}
and the helicity spinors are 
\begin{align}
{}_{\alpha}\SLH{p^I}=\left(\,\sqrt{E_p+|\vec p|}\left(\begin{array}{c} -s_{\theta/2}e^{-i\phi/2}\\ c_{\theta/2}e^{i\phi/2}
\end{array}\right)\,\,,\,\,\sqrt{E_p-|\vec p|}\left(\begin{array}{c} c_{\theta/2}e^{-i\phi/2}\\ s_{\theta/2}e^{i\phi/2}
\end{array}\right)\,\right)\\
{}^{\dot\alpha}\SRH{p^I}=\left(\,\sqrt{E_p-|\vec p|}\left(\begin{array}{c} -s_{\theta/2}e^{-i\phi/2}\\ c_{\theta/2}e^{i\phi/2}
\end{array}\right)\,\,,\,\,\sqrt{E_p+|\vec p|}\left(\begin{array}{c} c_{\theta/2} e^{-i\phi/2}\\ s_{\theta/2}e^{i\phi/2}
\end{array}\right)\,\right).
\end{align}
The following relations are inferred from the above definitions
\begin{align}
|p^I\rangle [p_I|&= \sigma \cdot p & 
|p_I]\langle p^I|&= \bar\sigma \cdot p\\
[p^I\,p^J]&= m \epsilon^{IJ} &
\langle p^I\,p^J\rangle&=-m\epsilon^{IJ}\\
({}_{\alpha}\SLH{p^I})^*&= {}_{\dot\alpha}\SRH{p_I}&
({}^{\dot\alpha}\SRH{p^I})^*&= -{}^{\alpha}\SLH{p_I}\end{align}
Finally we use the following prescription to turn an incoming particle into an outgoing one:
\begin{align}\label{InOut}
\SLH{-p^I}&=\SRH{p^I} & \SRH{-p^I}&=-\SRH{p^I},
\end{align}
so that $\SLH{-p}[-p|=-p\cdot\sigma$.
	
	\section{Loss of spin correlation}\label{SpLss}
As pointed out in Sec.~\ref{TDR}, the CP-violation effect encoded in spin correlations vanishes if the combined 
fragmentation and decay yields a diagonal structure in spin indices.
	The simplest example of the loss of spin information is hadronization to a pseudoscalar meson.  In this case 
eq.~(\ref{CGpS}) leads to 
	\begin{align}\nonumber
	\left({\mathcal A}_{q\to P_q\to 
		X}{\mathcal A}_{q\to P_q\to X}^\dagger\right)^I_J &=C
	\left(\langle q^I S_{\ell}^K\rangle+[S_\ell^K q^I]\right)\left([ S_{\ell,K} q_J]+\langle q_J S_{\ell,K} \rangle\right)\\
	&=2([q_J q^I]+\langle q^I q_J\rangle)=4 C m_q\delta^I_J
	\end{align}
	given that the spinor $S_\ell$ is built with the momentum of $q$.
	
	For the case of a vector meson decaying to a meson and a photon, as in the $B^*$ case, one has
	\begin{align} \nonumber
		\left({\mathcal A}_{q\to P_q^*\to 
		P_q\gamma^+}{\mathcal A}_{q\to P_q^*\to P_q\gamma^+}^\dagger\right)^I_J&=C [S_\ell^L\gamma][q^I \gamma]\langle\gamma S_{\ell,L}\rangle \langle \gamma q_J\rangle=C\langle \gamma q_J\rangle[q^I\gamma]\left(-2p_\gamma\cdot p_{P_q^*}\right),\\
	\left({\mathcal A}_{q\to P_q^*\to P_q\gamma^-}{\mathcal A}_{q\to P_q^*\to P_q\gamma^-}^\dagger\right)^I_J&=C \langle S_\ell^L\gamma\rangle\langle q^I \gamma\rangle [\gamma S_{\ell,L}] [ \gamma q_J]=C\langle \gamma q^I\rangle[q_J\gamma]\left(2p_\gamma\cdot p_{P_q^*}\right).
	\end{align}
	Each term preserves q-quark polarization, but if the experiment is insensitive to photon helicity the sum gives
	\begin{align}\nonumber
	\langle q^I p_\gamma q_J]-\langle q_J p_\gamma q^I]&=\mbox{Tr}\left(p_\gamma \varepsilon_{JJ'}\left(\SRH{q^K}\langle q^I|-\SRH{q^I}\langle q^K|)\right) \right)=\mbox{Tr}(p_{\gamma}\SLH{q_K}\langle q^K|)\delta^{I}_J\\&=2p_\gamma\cdot p_q\delta^{I}_J
	\end{align}
	
	
	The decay to a polarized photon has a spin resolution power of $1$ and the fragmentation into $P_q^*$ is 
$\approx 7$~\cite{Feindt:1995cf} greater than to $\Lambda_q$ baryons.  If such decays could be observed in a future 
experiment they would provide a promising avenue for a CP test of the $hbb$ interaction.
	
	The above two cases summarize the possible structures of a two-body decay of an excited hadron via the QCD or QED 
interaction.  Using only parity one can write in general 
	\begin{align}\nonumber
	\mathcal A_{q\to H}\left(\sum_{s_X,s_Y}d\Gamma_{H\to XY}\right)\mathcal A^\dagger_{q\to H}&= S (\langle q^I q_J\rangle-[q^I q_J])+J^\mu (\langle q^I \sigma^\mu q_J]-\langle q_J\sigma_\mu q^I])\\
	&=(2m_qS+2J\cdot p_q) \delta^I_J.
	\end{align} 
	To build the parity-even structure one can start from two incoming $q^I$ and $\bar q^J$ quarks as in eq.~(\ref{SpPV}) for 
Higgs-boson decay and flip the momentum of $\bar q$ to turn it into an outgoing quark, which gives the relative sign in eq.~(\ref{InOut}).
	

\section{Polarization resolution}\label{alPWR}
The spin resolution power $\alpha$ is determined using the following integrals:
\begin{align}
I_{\Lambda}=&\int \hat f_{1c}^2(x) \frac{(1-\delta +x)^2-4 x}{48} \left(2x+\delta-1 \right) dx\\
I_{\ell,in}=&\int\frac{y-x}{16} \left(1-\delta -\frac{2 x}{y}+x-y\right)\Theta(z)dxdy\\
I_{\ell,ex}=&\int\hat f_{1q}^2(x)\frac{y-x}{16} \left(1-\delta -\frac{2 x}{y}+x-y\right)\Theta(z)dxdy
\end{align}
with $z=1+x-\delta-x/y-y$, $\Theta$ the heaviside function, and the limits of integration 
in the first integral given by $x=0, (1-\sqrt{\delta})^2$.  To obtain the resolution 
power these integrals are divided by the appropriate integral, 
\begin{align}
I_{in}=&\int \frac{1}{16}(y-x)(1+x-\delta-y) \Theta(z)dx dy\\
I_{ex}=&\int \hat f_{1q}^2(x) \frac{1}{16}(y-x)(1+x-\delta-y)\Theta(z) dx dy.
\end{align}
In these integrals we model the semi-leptonic decay form factor using the Isgur-Wise function~\cite{Mannel1992},
\begin{align}
f(x)=\frac{\omega_0^2}{\omega_0^2+\frac{(1-\sqrt{\delta})^2-x}{\sqrt{\delta}}}
\end{align}
with $\omega_0=0.89$.  The spin resolution powers are given by 
\begin{align}
\alpha&=\frac{I_{\ell,in}}{I_{in}} &
\alpha&=\frac{I_{\ell,ex}}{I_{ex}} &
\alpha&=\frac{I_{\Lambda}}{I_{ex}} .
\end{align}

\bibliography{hqqCP_referee}
\bibliographystyle{JHEP}
\end{document}